\documentclass[twocolumn,amsmath,showkeys,amssymb,superscriptaddress,longbibliography,floatfix, pre]{revtex4-2}

\pdfoutput=1

\usepackage{bm}
\usepackage{graphicx,epsfig}
\usepackage{amsmath,amssymb,mathtools}
\usepackage{soul}
\usepackage{url}
\usepackage[rgb]{xcolor}
\usepackage{hyperref}
\def\be{\begin{equation}}
\def\ee{\end{equation}}

\def\bc{\begin{center}}
\def\ec{\end{center}}
\def\bea{\begin{eqnarray}}
\def\eea{\end{eqnarray}}
\newcommand{\avg}[1]{\langle{#1}\rangle}

\newcommand{\revise}[1]{{\color{black}#1}}

\begin{document}

\title{Triadic percolation on multilayer networks}

\author{Hanlin Sun}
\affiliation{Nordita, KTH Royal Institute of Technology and Stockholm University, Hannes Alfvéns väg 12, SE-106 91 Stockholm, Sweden}
\author{Filippo Radicchi}
\affiliation{Center for Complex Networks and Systems Research, Luddy School
  of Informatics, Computing, and Engineering, Indiana University, Bloomington, 
  47408, USA}
\author{Ginestra Bianconi}
\affiliation{School of Mathematical Sciences, Queen Mary University of London, London, E1 4NS, United Kingdom}

\begin{abstract}
Triadic interactions are special types of higher-order interactions that occur when regulator nodes modulate the interactions between other two or more nodes. In presence of triadic interactions, a percolation process occurring on a single-layer network becomes a fully-fledged dynamical system, characterized by period-doubling and a route to chaos. Here, we generalize the model to multilayer networks and name it as the multilayer triadic percolation (MTP) model. We find a much richer dynamical behavior of the MTP model than its single-layer counterpart. MTP displays a Neimark–Sacker bifurcation, leading to oscillations of arbitrarily large period or pseudo-periodic oscillations. Moreover, MTP admits period-two oscillations without negative regulatory interactions, whereas single-layer systems only display discontinuous hybrid transitions. This comprehensive model offers new insights on the importance of regulatory interactions in real-world systems such as brain networks, climate, and ecological systems.
\end{abstract}
\maketitle

\section{Introduction}

Many, if not all, real systems are more accurately represented in terms of higher-order rather than dyadic networks~\cite{bianconi2021higher,bick2023higher,battiston2021physics}.
Accounting for higher-order interactions is essential to properly understand critical phenomena emerging in these systems as the result of the complex interplay between network topology and dynamics~\cite{millan2025topology}; critical phenomena observed in higher-order networks display properties radically different from those observed when framed on dyadic networks~\cite{bianconi2021higher,bick2023higher,battiston2021physics,majhi2022dynamics}. 
Examples include synchronization 
\cite{millan2020explosive,carletti2023global,skardal2019abrupt,skardal2020higher,gambuzza2021stability,anwar2023synchronization},  epidemic spreading \cite{st2021universal, jhun2019simplicial,landry2020effect,iacopini2019simplicial}, percolation \cite{sun2021higher, bianconi2024theory}, and evolutionary dynamics \cite{alvarez2021evolutionary,civilini2021evolutionary, civilini2024explosive,sadekar2025drivers}.

Triadic interactions are general types of higher-order interactions describing the regulatory activity of nodes on interactions among other nodes~\cite{sun2023dynamic, sun2024higher,millan2024triadic, niedostatek2024mining, nicoletti2024information,millan2025spatio}. 
As a paradigmatic example, in brain networks, glia modulate the synaptic interactions between neuron pairs \cite{cho2016optogenetic} to form glia--neuron interactions,  
which are attracting great scientific interest lately~\cite{kozachkov2025neuron,kanakov2019astrocyte,menesse2025astrocyte,millan2025spatio,makovkin2021synchronization}. Another example of triadic interactions that can be observed in the brain are axo-axonic synapses, consisting of an axon terminating on another axon or axon terminal, which play a key role in presynaptic modulation \cite{cover2021axo, millan2025spatio}. Moreover, triadic interactions are also present in ecological networks \cite{bairey2016high,grilli2017higher}, and  biochemical reaction networks \cite{chen2023teasing}. 
Recently, an innovative statistical mechanics framework called {\em triadic percolation}~\cite{sun2023dynamic, sun2024higher,millan2024triadic}, combining  percolation theory and the theory of dynamical systems, has demonstrated that triadic interactions can trigger the  giant component to become time-dependent. 
Specifically,
the size of the giant component is characterized by spatio-temporal patterns 
potentially displaying 
period-doubling  and a route to chaos.  

Percolation \cite{li2021percolation, dorogovtsev2008critical, lee2018recent} is one of the most important critical phenomena defined on networks with wide applications ranging from assessing the robustness \cite{artime2024robustness} and fragility \cite{morone2015influence,gu2025deep} of real networks, to study  the distribution of entanglement in quantum networks \cite{meng2023percolation,hu2025unveiling,nokkala2024complex} and resource consumption in transportation and communication networks~\cite{kim2024shortest, meng2025path, kim2025modeling}. By predicting the relative size of the giant component (GC) after damage to nodes or links, percolation theory establishes the minimum prerequisite of network connectivity for dynamic processes to occur on the network. Additionally, percolation theory can be used to predict macroscopic network activity. Indeed, the fraction of nodes in the GC is commonly used as a proxy of the activity of network. It is known that the network structure strongly affect the critical behavior of percolation. On simple networks, node and link percolation displays a continuous second-order transition \cite{dorogovtsev2008critical}, while some generalized percolation problems, such as interdependent percolation on multiplex networks and higher-order percolation on hypergraphs, 
display discontinuous hybrid transitions  \cite{buldyrev2010catastrophic, bianconi2018multilayer,radicchi2017redundant,baxter2012avalanche,son2012percolation, dong2012percolation,sun2021higher}. In all these processes,  the GC, after an initial damage, 
can be affected by cascading failures, but eventually reaches a static stationary state that remains unchanged in time.

In real scenarios such as in brain dynamics or in climate, however, the 
GC
changes in time continuously, never reaching a static stationary state. Triadic percolation allows to properly capture the dynamic nature of percolation displayed by such complex systems \cite{sun2023dynamic, millan2024triadic, sun2024higher}. In triadic percolation, the links of a network are up- or down- regulated thanks to the presence of regulatory triadic interactions between nodes and links. Such a simple and intuitive ingredient turns percolation into a fully-fledged dynamical process displaying critical
properties that are very different
not only from those of a standard second-order phase transition as in ordinary percolation, but also from the ones that characterize the discontinuous hybrid phase transition of interdependent percolation\cite{buldyrev2010catastrophic,bianconi2018multilayer,baxter2012avalanche}. Indeed the phase diagram of triadic percolation \cite{sun2021higher} 
becomes
an orbit diagram, implying that the fraction of nodes in the 
GC
can display a bifurcation and a route to chaos. Moreover, on spatial networks, triadic percolation can generate dynamic topological patterns of the 
GC~\cite{millan2024triadic}.

Multilayer networks~\cite{bianconi2018multilayer, kivela2014multilayer, boccaletti2014structure} have been extensively explored in the past decade as a general framework for describing robustness and fragility of interconnected networks formed by nodes and links of different types, i.e., carrying a different functional role. For instance, the interaction between neurons and glia is a paradigmatic example of the multilayer network where glia are nodes of one layer and neurons are nodes of the other layer \cite{makovkin2021synchronization}.
Moreover, multilayer networks provide a very fertile ground for studying interactions among different regions of the brain \cite{reis2014avoiding}. So far, however, triadic percolation models have been investigated exclusively on single-layer networks and hypergraphs \cite{sun2023dynamic,sun2024higher,millan2025spatio}.

In this work, we propose a novel model of multilayer network with triadic interactions, called {Multilayer Triadic Percolation} (MTP) model, allowing for both {intralayer} and {interlayer} triadic regulatory interactions between nodes and links. We show that 
the simultaneous presence of both intralayer and interlayer triadic regulations induces novel dynamical states of the 
GC
that are not observed on single-layer networks with triadic interactions.
In particular, we observe, in addition to the period-doubling bifurcations and the route to chaos in the universality class of the logistic map as observed in single-layer networks, also new types of bifurcations. These include  the Neimark--Sacker bifurcation \cite{kuznetsov1998elements,ott2002chaos} from a steady state to a periodic or quasi-periodic oscillations of the size of the
GC
and the bifurcation leading to period-two oscillation between a silenced state (where the fraction of active nodes is zero) and an active state (where there is a non-zero fraction of active nodes). 
Our results highlight the rich behavior that may result from the combination of multilayer structural and regulatory interactions.


The paper is structured as follows. In Section \ref{sec:model}, we present our multilayer networks incorporating both intra- and inter-layer triadic regulations.  In Section \ref{sec:theory}, we define the multilayer triadic percolation (MTP) model. In Section \ref{sec:multiplexity}, we fully characterize the dynamics and the critical behavior of the MTP model; more specifically, we offer a comprehensive analysis of the critical behavior of the model in simplified scenarios (in the presence only of interlayer or intralayer triadic interactions) and in the general multilayer scenario. Finally, in Section \ref{sec:conclusion}, we provide our concluding remarks and future perspectives.

\section{Random multilayer networks with triadic interactions}\label{sec:model}

A triadic interaction occurs when a node regulates a structural edge between other two nodes.
In this work, in which we adopt the framework of triadic percolation already considered in Refs.~\cite{sun2023dynamic,sun2024higher,millan2025spatio}, we assume that this regulation implies switching on or off 
the edge
between the two considered nodes.

\begin{figure}
    \centering
    \includegraphics[width=0.95\linewidth]{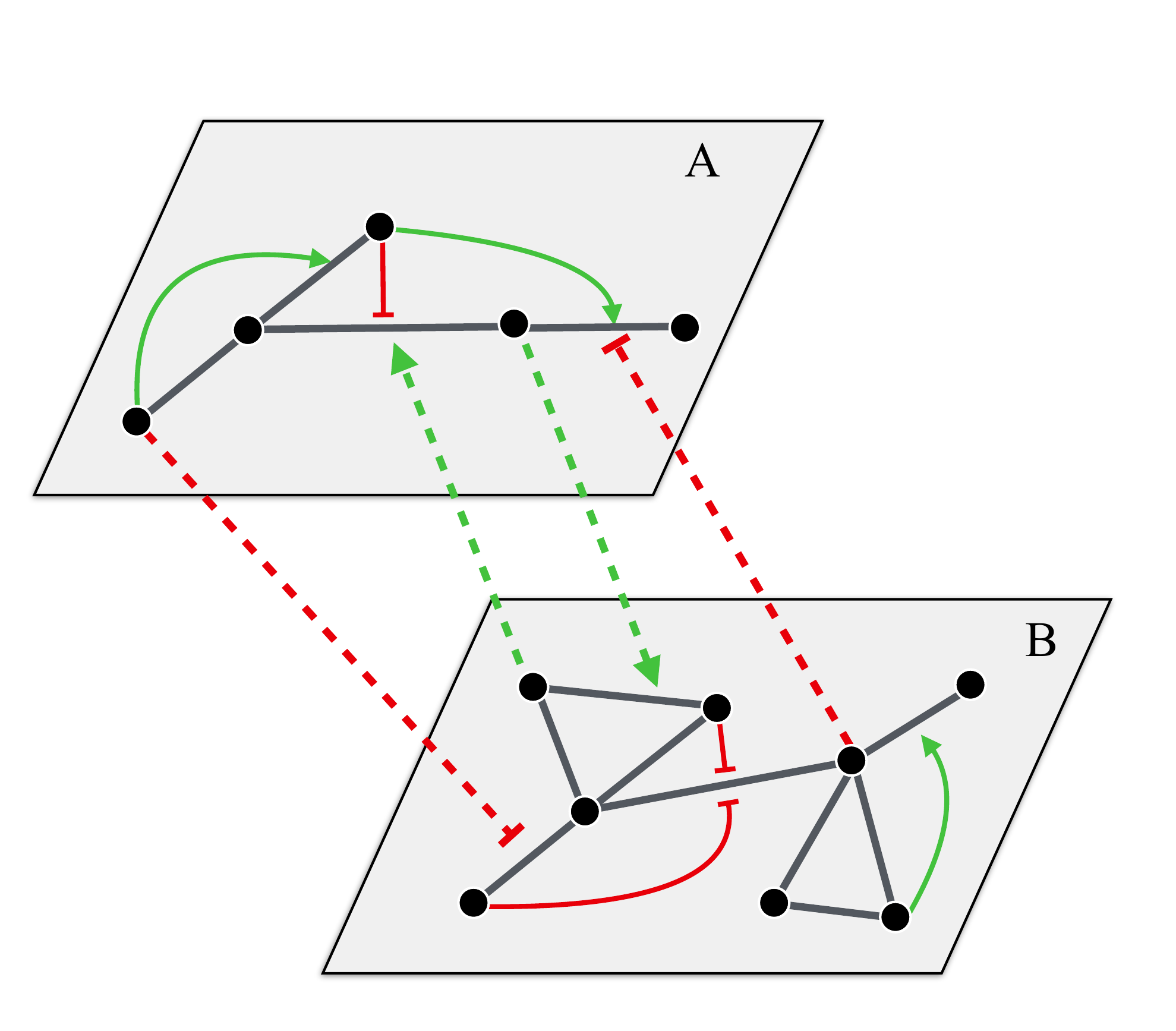}
    \caption{Schematic representation of a multilayer network with triadic regulatory interactions. 
    The network is composed of the two layers A and layer B.
    Nodes in the two layers are not one-to-one interdependent.
    Indeed in this example  layer A has $N_A=5$ nodes and layer B has $N_B = 8$ nodes. We distinguish two main types of interactions: 
    structural intralayer links (gray lines) between pairs of nodes within the same layer; 
    triadic regulatory interactions, either interlayer (dashed lines) or intralayer (solid lines), between regulator nodes and regulated structural links.
    A regulatory interaction can be either negative (red) or positive (green) depending on whether the regulator node down- or up-regulate the regulated structural link. 
    }
    \label{fig:diagram_regulation}
\end{figure}

We consider a multilayer network (for a schematic representation, see Fig.~\ref{fig:diagram_regulation}) with two layers, namely A and B, and with both interlayer and intralayer triadic interactions. The multilayer network is denoted as $\vec{G}=(\mathcal{G}_{A}, \mathcal{G}_{B}, \mathcal{W}_{AB}, \mathcal{W}_{BA})$ and is formed by two networks $\mathcal{G}_{A}$ and  $\mathcal{G}_B$ with intralayer structural links and triadic interactions, and the bipartite regulatory networks  $\mathcal{W}_{AB}$ and $\mathcal{W}_{BA}$ capturing the interlayer triadic interactions. Specifically, each layer $\mathcal{G}_{i}=(V_{i}, E_{i}, W_i)$ (with $i \in \{A, B\}$)  is formed not only by the node set $V_i$ of cardinality $|V_i|=N_i$ and the set $E_i$ of {\em structural links},  but also by the set $W_i$ that specifies the directed \textit{intralayer} signed regulatory interactions from regulator nodes in $V_i$ to structural links in $E_{i}$.  
The \textit{interlayer} interactions are captured instead by the bipartite networks $\mathcal{W}_{AB}$ (and $\mathcal{W}_{BA}$) which specify the directed {interlayer} signed regulatory interactions between nodes in layer $A$ and links in layer $B$ (nodes in layer $B$ and links in layer $A$). 
Note that the set of nodes in layer $A$ and in layer $B$ are not in a one-to-one correspondence.

 We define a generative model for random multilayer networks with triadic interactions.  In this model the structural network in layer $A$ (layer $B$) is a random network with given  (structural) degree distribution $P_A(k)$ ($P_B(k)$). Moreover, intralayer regulatory interactions are drawn at random while enforcing that each structural link in layer $A$ has a number of positive ($\hat{\kappa}^+$) and negative ($\hat{\kappa}^-$) regulatory interactions drawn from the distribution $\hat{P}_{A,intra}^\pm(\hat{\kappa}^\pm)$ where the regulatory nodes of each link are chosen uniformly at random among all the nodes of layer $A$. A similar construction is valid for the intralayer regulatory interactions in layer $B$ where each structural link has a number of positive ($\hat{\kappa}^+$) and negative ($\hat{\kappa}^-$) regulatory interactions drawn from the distribution $\hat{P}_{B,intra}^\pm(\hat{\kappa}^\pm)$. In an analogous way, the interlayer regulatory interactions are generated by enforcing that each structural link in layer $A$ (layer $B$) has a number $\hat{\kappa}^{\pm}$ of positive ($+$) or negative ($-$)  regulatory interactions drawn from the distribution $\hat{P}_{B, inter}^\pm(\hat{\kappa}^\pm)$ ($\hat{P}_{A, inter}^\pm(\hat{\kappa}^\pm)$). Note that the interlayer regulator nodes of each structural link in layer $A$ (layer $B$) are chosen uniformly at random from the nodes in layer $B$ (layer $A$).

For future 
convenience, 
let us define the generating function of above-mentioned distributions as
\bea
G_{0, A/B}(x) &=& \sum_{k} P_{A/B}(k) x^k, \nonumber\\ 
G_{1, A/B}(x) &=& \sum_{k} \frac{k P_{A/B}(k)}{\avg{k}} x^k,\\
G^\pm_{A/B, intra}(x) &=& \sum_{\hat{\kappa}^\pm} \hat{P}_{A/B, intra}^\pm(\hat{\kappa}^\pm) x^{\hat{\kappa}^\pm}, \nonumber\\
G^\pm_{A/B, inter}(x) &=& \sum_{\hat{\kappa}^\pm} \hat{P}_{A/B, inter}^\pm(\hat{\kappa}^\pm) x^{\hat{\kappa}^\pm}.
\eea
Moreover, let us denote the average degree of the distributions 
$P_{A/B}(k)$, $\hat{P}^\pm_{A/B, intra}$ and $\hat{P}^\pm_{A/B, inter}$ as $c_{A/B}$, $c^\pm_{A_{\text{intra}}/B_{\text{intra}}}$ and $c^\pm_{A_{\text{inter}}/B_{\text{inter}}}$, respectively.  \revise{As it was demonstrated in Ref. \cite{sun2023dynamic}, the dynamical properties of triadic percolation are only qualitatively affected by the choice of the degree distribution. Specifically, the nature of the bifurcation transition and the route to chaos does not depend on whether the degree distribution has a finite or divergent second moment as it happens in scale-free networks. These results demonstrated in \cite{sun2023dynamic} for triadic percolation on single layer higher-order networks are not expected to change for MTP. Therefore, in this work, without loss of generality,  only Poisson degree distributions are considered, thus the average degrees 
uniquely determine the mentioned distributions. }

\section{Multilayer triadic percolation (MTP)}\label{sec:theory}

We define the MTP model as a natural generalization of the triadic percolation model considered in Refs.~\cite{sun2023dynamic,sun2024higher,millan2025spatio}.
The novel ingredients here are that the network is composed of two layers and that structural edges are regulated simultaneously by intralayer and interlayer triadic regulatory interactions.  

The dynamics of the giant component (GC) in MTP is captured by the following 
iterative two-step algorithm. 
In Step 1 (percolation), we evaluate the GC of each structural network (layer A and layer B), and in each layer we consider as active the nodes that belong to the layer-wise GC.
In Step 2 (regulation),
we up- or down- regulate the structural links of each layer according to a Boolean rule which takes into account the activity of their interlayer and intralayer regulator nodes and stochastic noise.
Specifically, the MTP algorithm is defined as follows:



At time $t=0$, links in both layer $A$ and layer $B$  are randomly retained with probability $p^{(0)}_A$ (in layer A) and $p^{(B)}_B$ (in layer B). 
At any given time $t > 0$, the algorithm proceeds by iterating two steps:

\begin{itemize}
    \item Step 1. Given the configuration of activity of the structural links 
    at time $t - 1$, we determine the GC of each layer. 
Each node of layer $A$ (layer $B$) is considered  active if the node belongs to
the GC of the structural network of layer $A$ (layer $B$). 
The node is considered inactive otherwise.

    \item Step 2.  Given the set of all active nodes obtained at Step 1, 
    a link in layer $A$  (layer $B$) is active if   
    \begin{enumerate}
    \item[(a)] the link is regulated by at least one active positive regulator in both layers $A$ and $B$; 
    \item[(b)] the link is not regulated by any active negative regulator in either layer $A$ or $B$;
    \item[(c)] the link is not randomly deactivated with probability $1-p$.
    \end{enumerate}
\end{itemize}


Let us indicate with  $p_A^{(t)}$ ($p_B^{(t)}$) the probability that a link in layer $A$ (layer $B$) is retained at time $t$. In a random multilayer network with triadic interactions, Step 1 of MTP implements percolation \cite{bianconi2018multilayer}. Therefore, the fraction $R_{A}^{(t)}$ of nodes that at time $t$ are active, i.e. in the GC of layer $A$  is given by 
\bea
R_A^{(t)} &=& 1-G_{0,A}\left(1-S_A^{(t)} p_A^{(t-1)}\right), 
\eea
where $S_A^{(t)}$ is the probability that by following a link in layer $A$ we reach a node in the GC, which is known \cite{bianconi2018multilayer} to obey
\bea
S_A^{(t)} &=& 1-G_{1,A}\left(1-S_A^{(t)} p_A^{(t-1)}\right). 
\eea
Similarly, the fraction $R_{B}^{(t)}$ of active nodes in layer $B$  at time $t$ is dictated by the following two equations
\bea
R_B^{(t)} &=& 1-G_{0,B}\left(1-S_B^{(t)} p_B^{(t-1)}\right),\nonumber\\
S_B^{(t)} &=& 1-G_{1,B}\left(1-S_B^{(t)} p_B^{(t-1)}\right).
\eea
These equations express that at each time $t$ the fraction of active nodes $R_{A}^{(t)}$ ($R_{B}^{(t)}$) is uniquely determined by $p_A^{(t-1)}$ ($p_{B}^{(t-1)}$) thus we can define functions $f_A(x)$ and $f_B(x)$ such that 
\bea
R_A^{(t)} = f_A\left(p_A^{(t-1)}\right), \quad R_B^{(t)} = f_B\left(p_B^{(t-1)}\right).
\eea
Step 2 of the MTP algorithm determines the equations for $p_{A}^{(t)}$ and $p_B^{(t)}$ which encode the effect of the link regulation due to the presence of  signed interlayer and intralayer triadic interactions.
Specifically the MTP algorithm implies that $p_{A}^{(t)}$ and $p_B^{(t)}$ are given by 
\bea
p_A^{(t)} &=& p p_A^{intra}(t) p_A^{inter}(t), \nonumber\\
p_B^{(t)} &=& p p_B^{intra}(t) p_B^{inter}(t).
\label{pAB1}
\eea
where $p_{A/B}^{intra/inter}$ reflects the regulatory role of the intralayer and interlayer triadic interactions and are given by 
\bea
p_A^{intra}(t) &=& G_{A_{intra}}^{-}\left(1-R_A^{(t)}\right)\left[1-G_{A_{intra}}^{+}\left(1-R_A^{(t)}\right)\right], \nonumber\\
p_A^{inter}(t) &=& G_{B_{inter}}^{-}\left(1-R_B^{(t)}\right)\left[1-G_{B_{inter}}^{+}\left(1-R_B^{(t)}\right)\right], \nonumber\\
p_B^{intra}(t) &=& G_{B_{intra}}^{-}\left(1-R_B^{(t)}\right)\left[1-G_{B_{intra}}^{+}\left(1-R_B^{(t)}\right)\right], \nonumber\\
p_B^{inter}(t) &=& G_{A_{inter}}^{-}\left(1-R_A^{(t)}\right)\left[1-G_{A_{inter}}^{+}\left(1-R_A^{(t)}\right)\right].\nonumber
\label{pinterintra}
\eea
while $p$ denotes the probability that an up-regulated link is retained after random  due to stochastic noise. Therefore, the regulatory interactions that occur in Step 2 express how $p_{A}^{(t)}$ and $p_B^{(t)}$ depend on both $R_A^{(t)}$ and $R_B^{(t)}$.
It follows that both Step 1 and Step 2 can be encoded in the two-dimensional map:
\bea
p_A^{(t)}= g_A\left(R_A^{(t)}, R_B^{(t)}; p\right), \quad p_B^{(t)}= g_B\left(R_A^{(t)}, R_B^{(t)}; p\right),
\nonumber\\
R_A^{(t)} = f_A\left(p_A^{(t-1)}\right), \quad R_B^{(t)} = f_B\left(p_B^{(t-1)}\right),
\label{map}
\eea
where the functions $g_A(x,y)$ and $g_B(x,y)$ encode the dependency of $p_{A}^{(t)}$ and $p_{B}^{(t)}$ from $R_A^{(t)}$ and $R_B^{(t)}$ (Eq.~(\ref{pAB1}) and (\ref{pinterintra})).
MTP is described by a two-dimensional map because of the combined presence of both intralayer and interlayer triadic interactions.
In fact, single-layer triadic percolation in the absence of time delays and nested triadic interactions is captured only by a one-dimensional map.
This implies that, while for standard triadic percolation the size of the GC
undergoes period-doubling bifurcations and a route to chaos in the universality class of the logistic map \cite{sun2021higher}, for MTP, new dynamical and critical phenomena can in principle occur  
as the result of the network being composed of two layers. \revise{Note that here we focus on multilayer networks comprising two layers. More in general for multilayer networks with $M$ layers $\{A,B,C,\ldots\}$, triadic percolation is encoded in a $M$-dimensional map with $M>2$ that generalizes the two dimensional map defined in  Eq.~(\ref{map}), as follows
\bea
p_i^{(t)} &=& g_i\left(R_A^{(t)}, R_B^{(t)}, R_C^{(t)},   \ldots ;
p\right), \nonumber\\
R_i^{(t)} &=& f_i\left(p_i^{(t-1)}\right), \quad i=A, B, C, \ldots 
\eea
The higher-dimensional maps with $M>2$ are known to have a complex dynamical behavior which strongly affects their route to chaos \cite{ott2002chaos}. However, in this work our focus will be mostly the characterization of the local bifurcations from steady states and the critical properties of the MTP model. For general higher-dimensional maps of order $M\geq 2$, the critical local instability leads to one of the following three scenarios: discontinuous transitions, period-two oscillations and Neimark-Sacker bifurcations. The discussion  conducted here for a multilayer networks with $M=2$ layers can be directly extended, with the necessary precautions, to the more general case of multilayer networks with $M>2$.}
\section{Interplay between 
network multilayer structure and critical behavior}
\label{sec:multiplexity}
In this section, our goal is to demonstrate how 
multilayer regulatory interactions
affect the critical behavior of MTP.
Specifically, we investigate the transition between a phase with static GC and a phase with dynamic GC occurring at a critical value $p_c$ of the occupation probability $p$ of the structural links.
Given the deep connections between MTP and dynamical systems, captured by Eq.~(\ref{map}), this transition is characterized by the bifurcation of the two-dimensional map capturing the dynamics of triadic percolation.

We distinguish
three cases: (i) triadic interactions that are exclusively intralayer,  (ii) triadic interactions that are both intralayer and interlayer, (iii) triadic interactions that are exclusively interlayer.

\subsection{Triadic interactions that are exclusively intralayer}
\begin{figure}
    \centering
    \includegraphics[width=\linewidth]{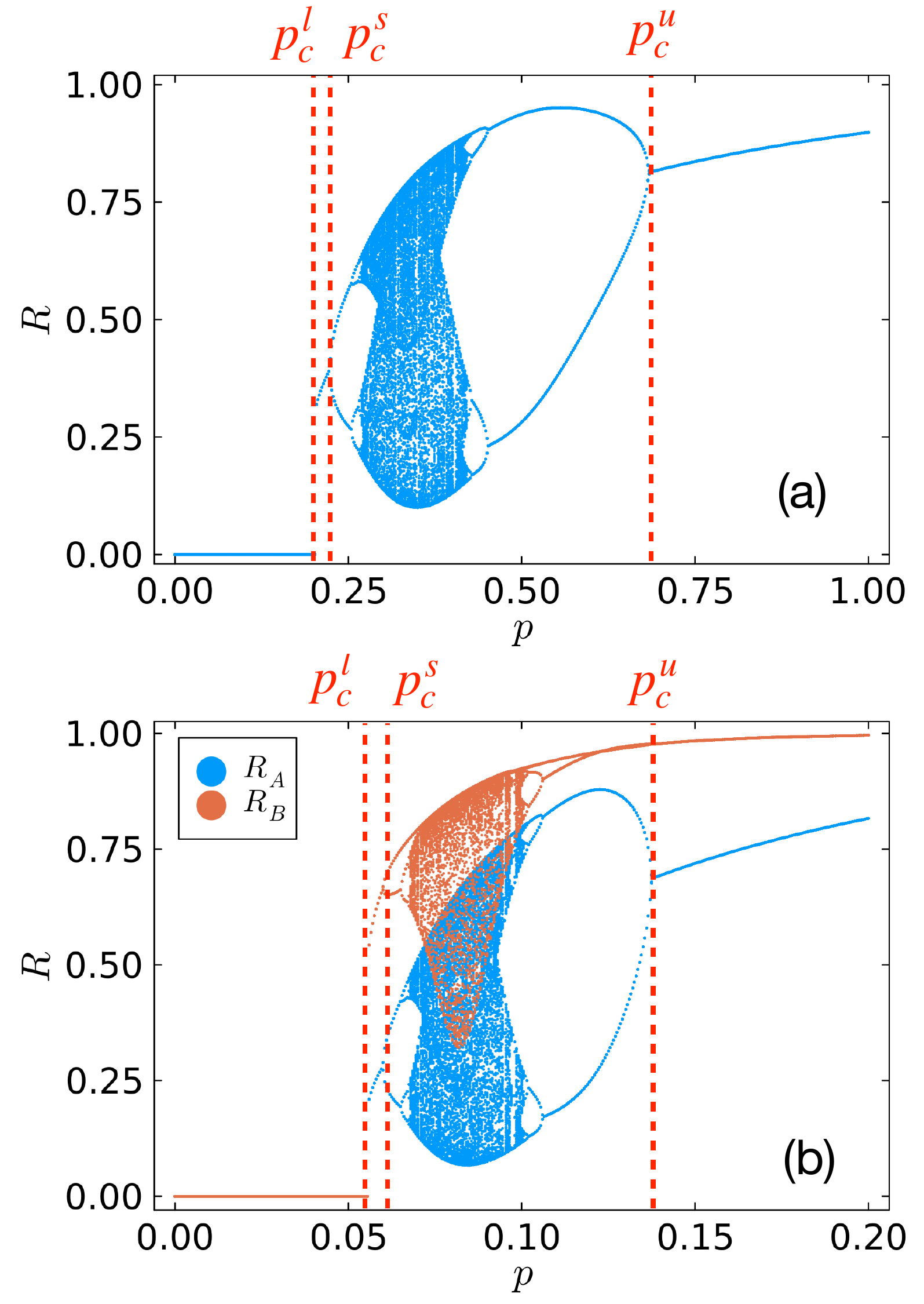}
    \caption{Orbit diagrams of triadic percolation on single-layer networks (a) and multilayer networks (b). We characterize the orbit diagram via upper stability threshold $p_c^u$, re-stabilization threshold $p_c^s$ and lower stability threshold $p_c^l$. In panel (a), the model parameters are $c=20$, $c^+=1.8$, $c^-=2.5$. In panel (b), the model parameters are $c_A=c_B=30$,  $c_{A_\text{inter}}^{+} = 10$,  $c_{A_\text{intra}}^{+} = 10$, $c_{A_\text{inter}}^{-} = 0.1$, $c_{A_\text{intra}}^{-} = 1.3$,  
    $c_{B_\text{inter}}^{+} = 20$, $c_{B_\text{intra}}^{+} = \infty$, 
     $c_{B_\text{inter}}^{-} = 0$,  $c_{B_\text{intra}}^{-} = 0$.}

    \label{fig:orbit}
\end{figure}

When triadic interactions are exclusively intralayer, we need to consider the model in which the probabilities $p_A^{(t)}$ and $p_B^{(t)}$ are given by 
\bea
p_A^{(t)}= p p_A^{intra}(t), \quad
p_B^{(t)}= p p_B^{intra}(t) ,
\eea
meaning that the two layers are completely decoupled. The dynamics on each layer can be simply described by a one-dimensional map
\bea
p_{A/B}^{(t)}= g_{A/B}\left(R_{A/B}^{(t)};p\right), \quad R_{A/B}^{(t)} = f_{A/B}\left(p_{A/B}^{(t-1)}\right).
\eea
Hence, in each layer, the scenario reduces to triadic percolation taking place on a single-layer network, which has been studied in Ref. \cite{sun2023dynamic}. In this scenario, we can study each layer independently. 
Thus, in each layer, starting from a stationary state, we can only observe one of two scenarios: a period-doubling bifurcation that gives rise to oscillatory behavior, or a discontinuous hybrid transition in which the non-zero giant component collapses to a vanishing giant component.
These two distinct scenarios can be predicted by considering the one-dimensional maps:
\bea
R_A^{(t)} &=& f_A\left(g_A\left(R_A^{(t-1)}; p\right)\right) := h_A\left(R_A^{(t-1)}; p\right) \nonumber\\
R_B^{(t)} &=& f_B\left(g_B\left(R_B^{(t-1)}; p\right)\right) := h_B\left(R_B^{(t-1)}; p\right). 
\eea
The stationary states $R_A^{\star}$ and $R_B^{\star}$ obeying 
\bea
R_A^{\star}=h_A(R_A^{\star}),\quad R_B^{\star}=h_B(R_B^{\star}),
\eea
lose stability for $J_A=h_A^{\prime}(R_A^{\star})=\pm 1$ and $J_B=h_B^{\prime}(R_B^{\star})=\pm 1$.
Specifically, the values $J_{A/B}=1$ occur for values of $p$ at which we observe a discontinuous transition, while the values $J_{A/B}=-1$ correspond to values of $p$ at which we observe the period-doubling bifurcation. These bifurcations denote the instability of the stationary order parameter of triadic percolation in layer A or in layer B. A rigorous analysis of the one-dimensional maps reveals that this is the onset of a route to chaos in the universality class of the logistic map.
If we focus on the instability of the stable solution of triadic percolation in single-layer networks, we can observe that, interestingly,  the period-doubling bifurcation transition can only occur in the presence of negative regulations \cite{sun2023dynamic}. Note that the period-doubling bifurcation can occur for different values of $p$ in the same network topology and that these bifurcations can also occur in networks in which the discontinuous hybrid transition is present. Figure \ref{fig:orbit} (a) shows a typical example of phase diagram of triadic percolation on single-layer networks that exhibits a route to chaos in the universality class of the logistic map and three critical points where the non-trivial stationary state changes stability, namely an upper bifurcation threshold $p_c^u$, a re-stabilization threshold $p_c^s$, and a lower bifurcation threshold $p_c^l$. At $p=1$, the dynamics typically converges to a stationary state. As the control parameter $p$ decreases, at the upper bifurcation threshold $p_c^u$, the fixed point loses stability and bifurcates. As $p$ continues decreasing, typically the dynamics undergoes period-doubling and route to chaos. Eventually, at the re-stabilization threshold $p_c^s$, the different branches of the cycles or chaos merge, and a re-stabilized fixed point emerges. When continuing decreasing $p$, the re-stabilized fixed point loses stability at $p_c^l$ and the trivial fixed point with null GC becomes the only stable fixed point, which the dynamical system converges to.
As shown in Figure $\ref{fig:orbit} (b)$, MTP with both intra- and inter-layer regulation also exhibits a route to chaos. However,  the mechanisms by which the stable solution loses stability can correspond to different types of critical behavior with respect to triadic percolation in single layers, possibly indicating a different universality class for the route to chaos as well.

\begin{figure*}
    \centering
    \includegraphics[width=\linewidth]{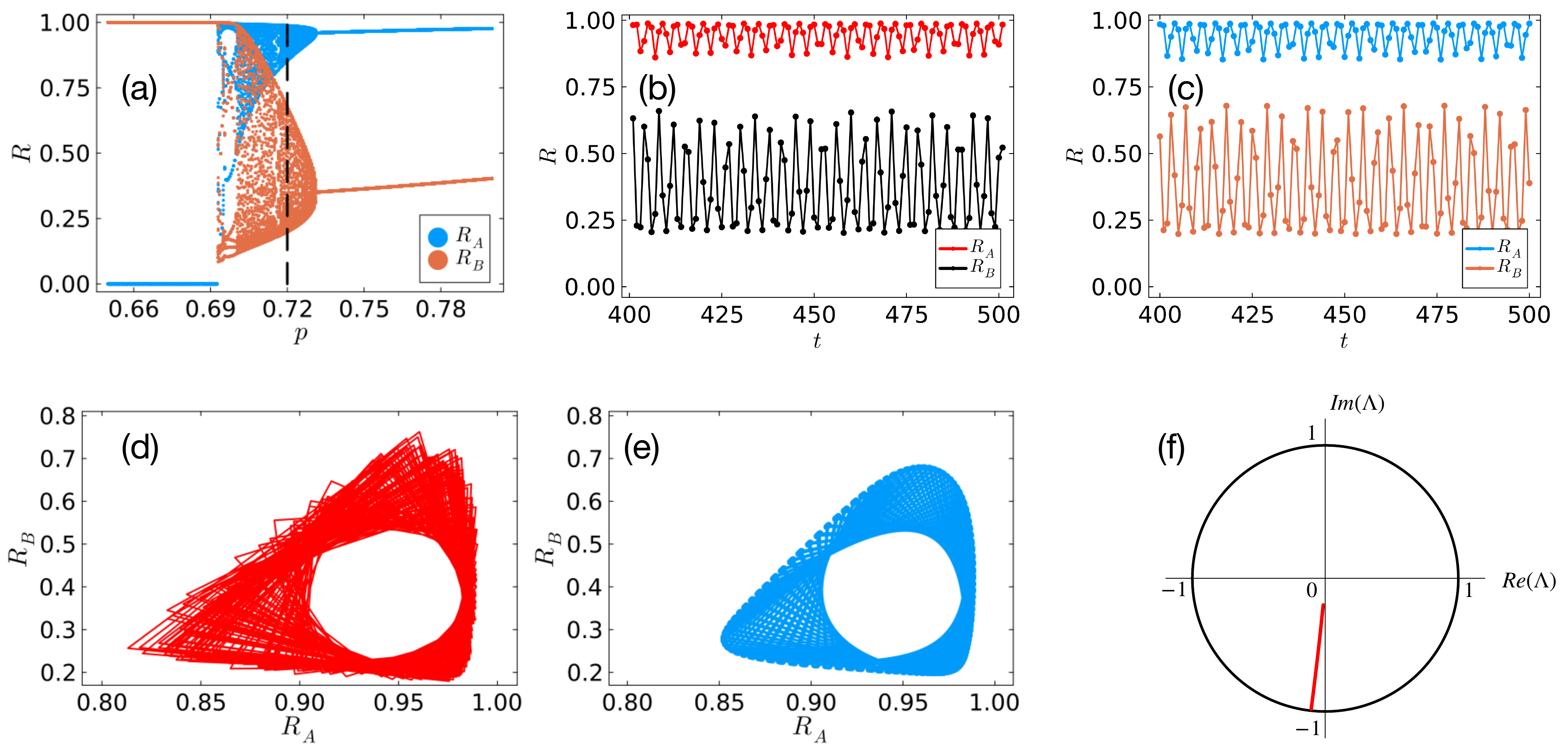}
    \caption{Neimark--Sacker bifurcation of multilayer triadic percolation with the presence of both intralayer and interlayer regulations. (a) Orbit diagram of the order parameters $R_A$ and $R_B$. (b) Monte Carlo simulation of the time series of the dynamics at $p=0.72$. The corresponding value is indicated by the black dashed line in panel (a). (c) Theoretical time series of the dynamics at the same $p=0.72$. (d) Monte Carlo simulation of the time evolution of $(R_A, R_B)$ at $p=0.72$. A spiral periodic (quasi-periodic) orbit is shown. (e) Theoretical time evolution of $(R_A, R_B)$ at $p=0.72$. (f) The leading eigenvalue $\Lambda$ of the Jacobian evaluated at the fixed point $(R_A^\star, R_B^\star)$ (red line). The eigenvalues cross the unit circle transversely, signalling the onset of a Neimark–Sacker bifurcation.
    The model parameters are $c_{A_\text{intra}}^{+} = 5$, $c_{A_\text{intra}}^{-} = 1.5$, $c_{A_\text{inter}}^{+} = \infty$, $c_{A_\text{inter}}^{-} = 3$, $c_{B_\text{intra}}^{+} = \infty$, $c_{B_\text{intra}}^{-} = 0$, $c_{B_\text{inter}}^{+} = 3$, $c_{B_\text{inter}}^{-} = 0$. In panel (b), the Monte Carlo simulation is conducted on a quenched network with \revise{$N_A=N_B=5 \times 10^5$ nodes.}}
    \label{fig:NS}
\end{figure*}

\begin{figure*}
    \centering
    \includegraphics[width=\linewidth]{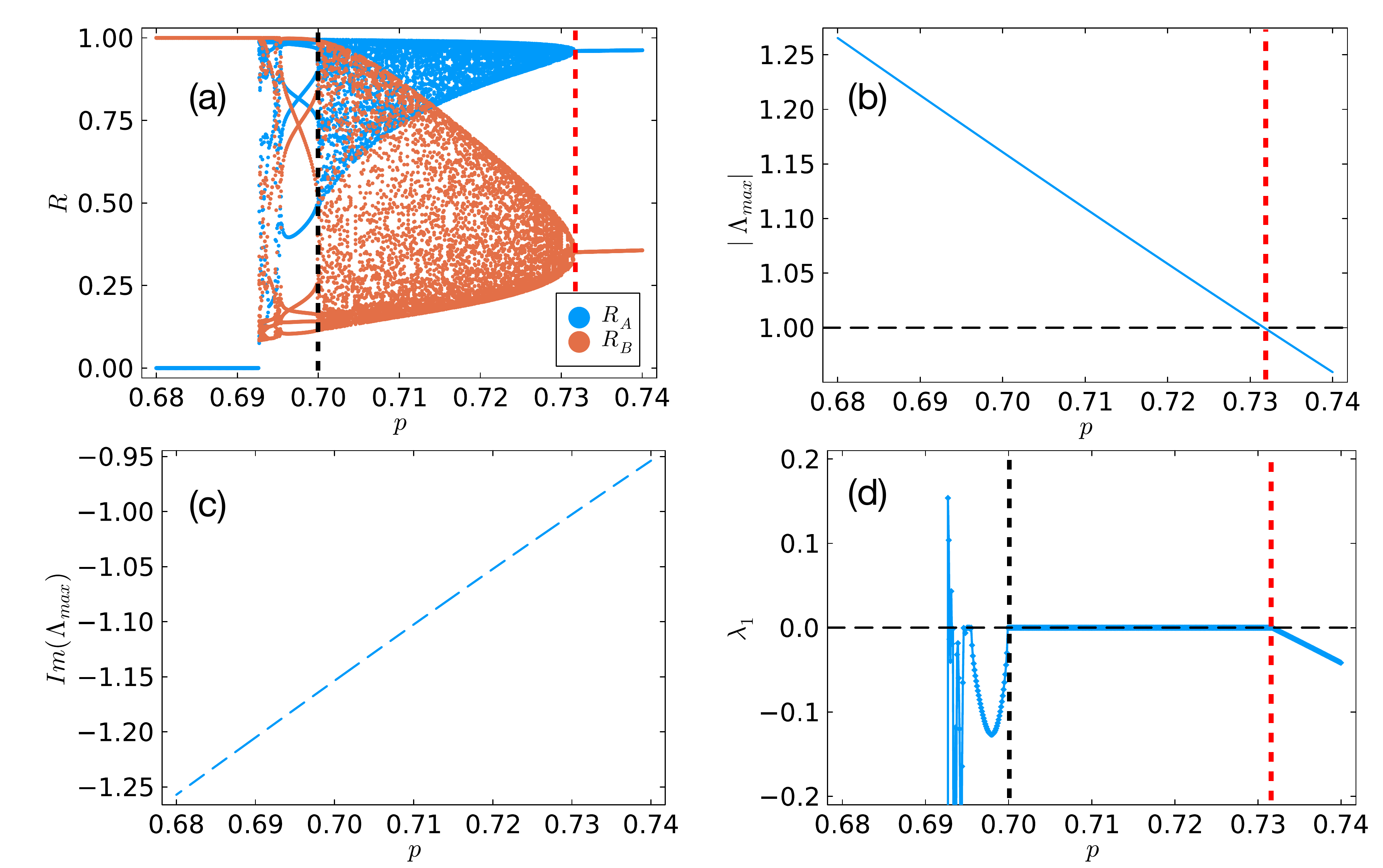}
    \caption{Characterization of the Neimark-Sacker bifurcation. In panel (a) we show a zoomed orbit diagram generated using the same parameters as Figure \ref{fig:NS} (a). The red and black vertical dashed line indicates the onset of pseudo-periodic oscillation via Neimark-Sacker bifurcation and the onset of periodic oscillaation via other types of bifurcations. When the pseudo-periodic oscillation emerges, the leading eigenvalue of the Jacobian defined in Eq. \ref{JNS} has modulus one (b) and non-zero imaginary part (c). In panel (d) we show the largest Lyapunov exponent $\lambda_1$ as a function of $p$.}
    \label{fig:NS_Lyapunov}
\end{figure*}

\begin{figure*}
    \centering
    \includegraphics[width=\linewidth]{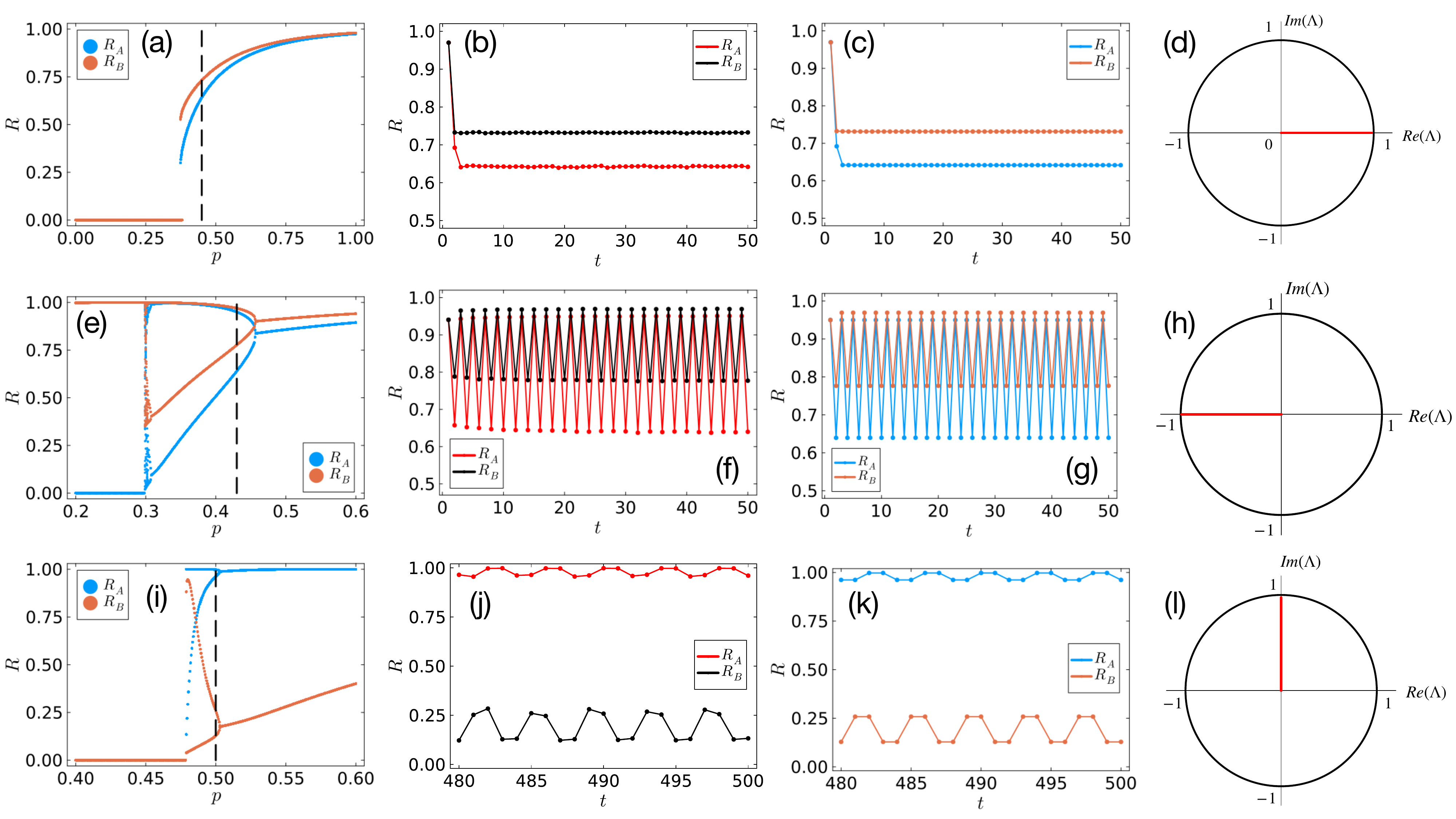}
    \caption{Different types of bifurcation of triadic percolation on multilayer networks with exclusively interlayer triadic interactions. We show three different bifurcations: (a)-(d) discontinuous transition, (e)-(h) period-doubling transition, and (i)-(l) Neimark--Sacker bifurcation. In the second and third columns, we show the Monte Carlo simulation (the second row) and theory result (the third row) of the time series of the dynamics at $p=0.45$ (panel (b), (c)), $p=0.43$ (panel (f), (g)), and $p=0.50$ (panel (j), (k)). The corresponding $p$ values are indicated as the black dashed line in the first column. In the fourth column, we plot the leading eigenvalue $\Lambda$ of the Jacobian evaluated at the fixed point $(R_A^\star, R_B^\star)$ (red line). 
    The eigenvalue crosses the unit circle at different angles indicating different natures of bifurcations: a discontinuous transition (panel (d)), a period-doubling bifurcation (panel (h)), and a special Neimark–Sacker bifurcation, where the eigenvalues form a purely imaginary pair (panel (l)). 
    The model parameters are summarized as follows. The structural and regulatory networks have both Poisson degree distributions. In the first row, only interlayer regulations are considered, the parameters are $c_{A_\text{inter}}^{+} = 3$, $c_{A_\text{inter}}^{-} = 0$, $c_{B_\text{inter}}^{+} = 10$, $c_{B_\text{inter}}^{-} = 0$. In the second row, the parameters are $c_{A_\text{intra}}^{+} = 10$, $c_{A_\text{intra}}^{-} = 2.2$, $c_{A_\text{inter}}^{+} = \infty$, $c_{A_\text{inter}}^{-} = 2$, $c_{B_\text{intra}}^{+} = \infty$, $c_{B_\text{intra}}^{-} = 0$, $c_{B_\text{inter}}^{+} = 10$, $c_{B_\text{inter}}^{-} = 0$. 
    In the third row, only interlayer regulations are considered, the parameters are $c_{A_\text{inter}}^{+} = 2$, $c_{A_\text{inter}}^{-} = 2.5$, $c_{B_\text{inter}}^{+} = 2$, $c_{B_\text{inter}}^{-} = 0$. 
    All Monte Carlo simulations are conducted on quenched networks of size \revise{$N_A=N_B=5 \times 10^5$ nodes.}}
    \label{fig:d_p2_p4}
\end{figure*}
\subsection{Triadic interactions that are both interlayer and intralayer}

In the presence of both intralayer and interlayer triadic interactions, we observe a dynamics of the MTP model that cannot be simply reduced to the dynamics observed in triadic percolation for single-layer networks. This scenario is indeed very distinct as the dynamics is captured by a two-dimensional map~\cite{ott2002chaos} rather than a one-dimensional map. In this case, as well we observe a route to chaos (see Figure $\ref{fig:orbit}$). Leaving the discussion about the universality class to later works, here we focus on the characterization of the bifurcation transitions. Specifically, we demonstrate that MTP   displays different types of bifurcation transitions with respect to single-layer networks
leading to new critical phenomena in the context of generalized percolation transitions.
The two-dimensional map that describes MTP in the presence of both interlayer and intralayer triadic interactions is given by
\bea
R_A^{(t)} &=& f_A\!\left( g_A\!\left( R_A^{(t-1)}, R_B^{(t-1)}; p \right) \right), \nonumber\\
R_B^{(t)} &=& f_B\!\left( g_B\!\left( R_A^{(t-1)}, R_B^{(t-1)}; p \right) \right).
\label{2map}
\eea
which can be written in a more compact way as
\bea
R_A^{(t)} &:=& h_A\!\left( R_A^{(t-1)}, R_B^{(t-1)}; p \right), \nonumber\\
R_B^{(t)} &:=& h_B\!\left( R_A^{(t-1)}, R_B^{(t-1)}; p \right).
\label{2hmap}
\eea
The stationary states $R_A^{\star}$ and $R_B^{\star}$ obey the stationary equation 
\bea
R_A^{\star}=h_A(R_A^{\star},R_B^{\star}),\quad R_B^{\star}=h_B(R_A^{\star},R_B^{\star}).
\label{Rstar}
\eea
The bifurcation points at which these stationary states lose stability can be predicted by considering the Jacobian $J$ of the two-dimensional map:
\bea
J\left[R_A^{\star}, R_B^{\star}\right] = \left.\begin{bmatrix}
\frac{\partial h_A}{\partial R_A} & \frac{\partial h_A}{\partial R_B} \\
\frac{\partial h_B}{\partial R_A} & \frac{\partial h_B}{\partial R_B}
\end{bmatrix}\right|_{R_A=R_A^{\star}, R_B=R_B^{\star}}
\label{JNS}
\eea
Since the eigenvalues of the Jacobian must be roots of a quadratic equation, there are only two possible scenarios: either both eigenvalues are real, or both eigenvalues are complex conjugate pairs. Due to the spectrum properties of the Jacobian, interesting critical phenomena can be observed. In particular, for certain choices of parameters, the Neimark--Sacker bifurcation occurs \cite{kuznetsov1998elements,ott2002chaos}. In this case,  the fixed point loses its stability and an invariant cycle emerges with an arbitrary period, or displays a quasi-periodic dynamics. 

In order to illustrate the emergence of the Neimark--Sacker transition in MTP, let us write the  Jacobian $J$ defined in Eq.(\ref{JNS}) as
\bea
J\left[R_A^{\star}, R_B^{\star}\right] = \begin{bmatrix}
A & B\\
C & D
\end{bmatrix}.
\eea
The eigenvalues of the Jacobian $J$ can be expressed as 
\bea \Lambda_{\pm} = \frac{\tau \pm \sqrt{\tau^2-4\Delta}}{2} \eea where
$\tau=A+D$ and $\Delta = AD-BC$. Depending on their values, a discontinuous hybrid transition, a period-doubling bifurcation, or a Neimark--Sacker bifurcation  \cite{kuznetsov1998elements} can be observed. Let us discuss the conditions corresponding to these critical phenomena.

Let us denote the eigenvalues as $\Lambda_+$ and $\Lambda_-$. When both eigenvalues are real, i.e., $\tau^2-4 \Delta > 0$, let us assume that $|\Lambda_{+}| > |\Lambda_{-}|$. If $\Lambda_+=1$ at criticality, a discontinuous hybrid transition can be observed (see Figure \ref{fig:d_p2_p4} (a-d)). If $\Lambda_+=-1$ at criticality, a period-doubling bifurcation can be observed (see Figure \ref{fig:d_p2_p4} (e-h)). On the other hand, when the eigenvalues are complex conjugate pairs, which occurs when $\tau^2-4 \Delta< 0$, a Neimark--Sacker bifurcation occurs when the modulus of the complex eigenvalues $|\Lambda_+|=|\Lambda_-|=1$ (see Figure \ref{fig:NS}). At the critical point, we denote
\bea
\Lambda =\Lambda_+= e^{ i \theta}
\eea
where $\theta$ is the rotation angle of the linearized map. Thus, the oscillation is periodic or quasi-periodic, depending on the value of $\theta$. Specifically, if $\theta$ is a rational multiple $k=\hat{p}/\hat{q}$ of $2\pi$, where $\hat{p}, \hat{q} \in \mathbb{N}$, the oscillation has period-$\hat{q}$. Otherwise, the oscillation is quasi-periodic.

Note that the critical phenomenon of Neimark--Sacker bifurcation arises from the complex eigenvalues of the Jacobian and thus cannot occur in one-dimensional systems, i.e., triadic percolation on single-layer networks. Second, the presence of the general Neimark--Sacker bifurcation requires (i) both positive and negative regulatory interactions and (ii) both interlayer and intralayer regulatory interactions. If the regulations are exclusively positive, the Jacobian has only 
positive
entries. Therefore, according to the Perron-Frobenius theorem, the leading eigenvalue is real. On the other hand, if the regulations are exclusively interlayer, the Jacobian has only zero entries on the diagonal, $A=D=0$, hence the eigenvalues are either real or purely imaginary, which is the case we will discuss in more detail in Sec. \ref{sec:exclusive_interlayer}. 

In Figure \ref{fig:NS}, we present the theoretical orbit diagram of the Neimark–Sacker bifurcation together with theoretical time series within the supercritical regime of the bifurcation. The theoretical results are validated by extensive Monte Carlo simulations on quenched networks (Figure \ref{fig:NS} (a-c)).
The Neimark–Sacker bifurcation occurs when a pair of complex-conjugate eigenvalues of the Jacobian at a fixed point cross the unit circle in the complex plane at a non-trivial angle, i.e., neither on the real nor imaginary axis. (see Figure \ref{fig:NS} (f)). At this critical parameter value, the fixed point loses stability and a closed invariant curve emerges.
The resulting dynamics are characterized by periodic or quasi-periodic oscillations on the invariant curve, depending on the ratio of the angular frequencies involved. Geometrically, in the phase plane of the order parameters $(R_A, R_B)$, this manifests as a spiral trajectory converging toward the invariant cycle (see Figure \ref{fig:NS} (d-e)). 
Moreover, the complex interplay between layers induces interesting responses of network activity when changing the strength of random damage. Different from the single-layer counterpart, increasing the strength of random damage (smaller $p$ value) can induce a higher \textit{steady} network activity (see Figure \ref{fig:NS} (a)). Interestingly, this theoretical prediction captures the overall behavior also of extensive numerical simulations of MTP. We expect, on the basis of the self-averaging properties of percolation, that the deviations found in the simulations are a finite size effect and will be suppressed in the large network limit.

\revise{To further characterize the bifurcation, in Figure \ref{fig:NS_Lyapunov} we show the leading eigenvalue $\Lambda_{max}$ of the Jacobian and the largest Lyapunov exponent $\lambda_1$ as functions of the control parameter $p$. The onset of the Neimark--Sacker instability is signalled by a complex-conjugate multiplier crossing the unit circle, i.e., $|\Lambda_{\max}|=1$ while $\mathrm{Im}(\Lambda_{\max})\neq 0$ (Figure \ref{fig:NS_Lyapunov} (b, c)). Consistently, the largest Lyapunov exponent increases from negative values that indicate stable fixed points, to values close to zero at the bifurcation point, and remains approximately zero throughout the regime where quasi-periodicity is present. This behavior indicates the emergence of invariant cycles and is consistent with the supercritical Neimark–Sacker bifurcation. Upon further decreasing $p$, the invariant curve may undergo additional bifurcations, leading to periodic windows and chaotic dynamics, which are respectively characterized by $\lambda_1<0$ and $\lambda_1>0$ (Figure \ref{fig:NS_Lyapunov} (d)).}

\begin{figure*}[!htb!]
    \centering
    \includegraphics[width=0.9\linewidth]{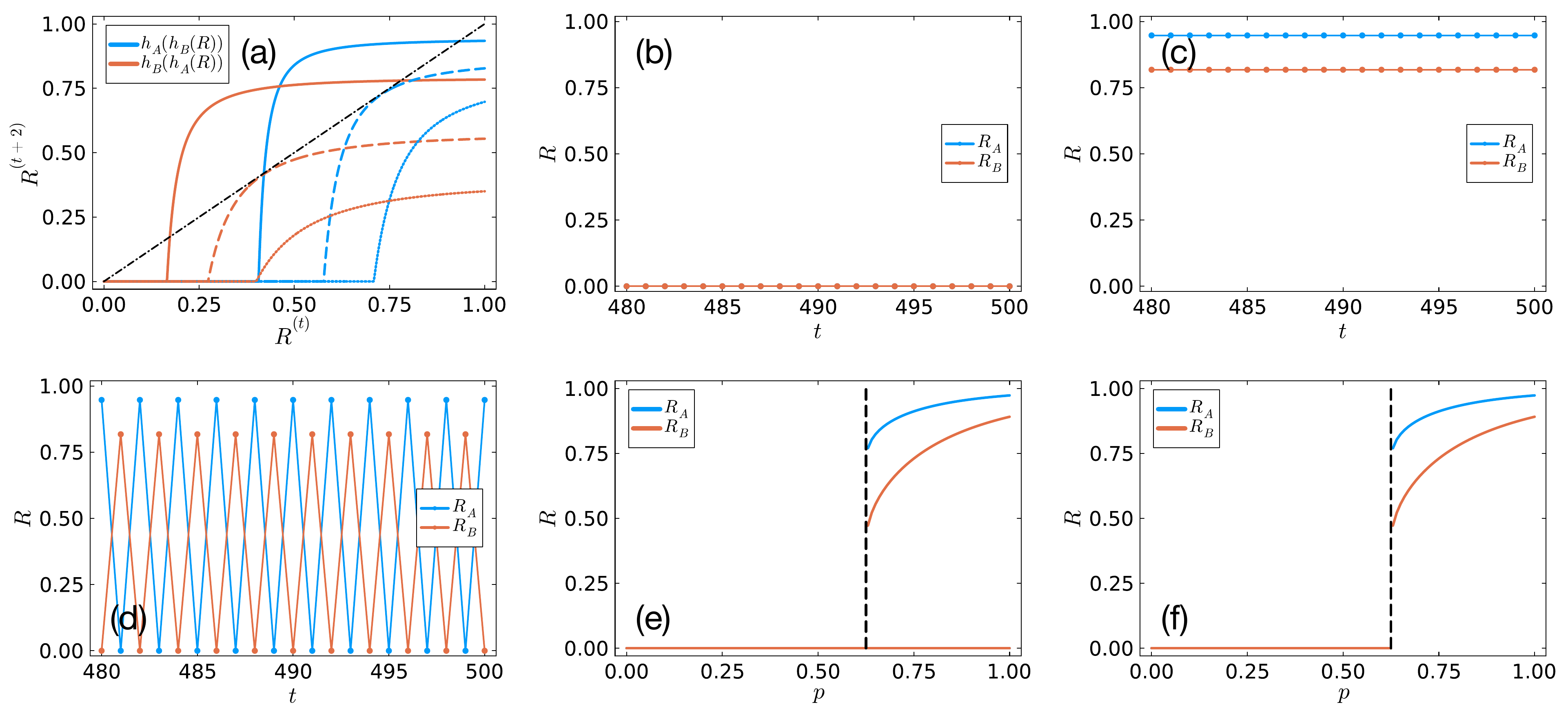}
    \caption{Dynamics and orbit diagrams of triadic percolation on multilayer networks with exclusively interlayer regulations. (a) The iterative map defined in Eqs. \ref{eq:second_iteration_map} at $p=0.85$ (solid line), $p=0.625$ (dashed line) and $p=0.55$ (dotted line). At the critical value $p=0.625$, a non-trivial stable fixed point emerges. (b-d) The time series of the dynamics at $p=0.85$, with initial conditions $p_A^0=p_B^0=0.02$ (b), $p_A^0=p_B^0=0.9$ (c), and $p_A^0=0.02, p_B^0=0.9$ (d). In panels (e-f), we show the orbit diagram of the dynamics with initial conditions $p_A^0=0.02, p_B^0=0.9$ (e) and $p_A^0=p_B^0=0.9$ (f). The parameters used here are $c_A=c_B=4$, $c_{A_{\text{inter}}}^+=1$, $c_{B_{\text{inter}}}^+=3$, $c_{A_{\text{inter}}}^- = c_{B_{\text{inter}}}^-=0$.}
    \label{fig:period2}
\end{figure*}


\subsection{Triadic interactions that are exclusively interlayer}\label{sec:exclusive_interlayer}
If the triadic interactions are exclusively interlayer, the two-dimensional map describing the MTP model given in general by Eq.(\ref{2hmap}) simplifies to
\bea
R_A^{(t)} =  h_A\left(R_B^{(t-1)}; p\right), \quad
R_B^{(t)} =  h_B\left(R_A^{(t-1)}; p\right),
\label{map_inter}
\eea
while the fixed point Eq.(\ref{Rstar}) reads 
\bea
R_A^{\star} =  h_A\left(R_B^{\star}; p\right), \quad
R_B^{\star} =  h_B\left(R_A^{\star}; p\right).
\eea
The stability of the fixed point can be investigated by studying the spectrum of the Jacobian of the map, which acquires in this case the simplified expression
\bea
J\left[R_A^{\star}, R_B^{\star}\right] &=& \left.\begin{bmatrix}
0 & \frac{\partial h_A}{\partial R_B} \\
\frac{\partial h_B}{\partial R_A} & 0
\end{bmatrix}\right|_{R_A=R_A^{\star}, R_B=R_B^{\star}}.
\label{eq:2d_jacobian}
\eea
The eigenvalues of the Jacobian are therefore given by
\bea
\Lambda_{\pm}[J] = \pm \sqrt{\frac{\partial h_A}{\partial R_B} \frac{\partial h_B}{\partial R_A}}. 
\eea

There are three possible ways in which  the fixed points $(R_A^{\star}, R_B^{\star})$ can lose stability. When $\frac{\partial h_A}{\partial R_B} \frac{\partial h_B}{\partial R_A} > 0$, the eigenvalues are real, hence the fixed points lose stability at  ${\Lambda}=\max(\Lambda_{+},\Lambda_{-}) =  1$, leading to discontinuous hybrid transitions or period-two oscillations depending on the values of the initial conditions, as we will discuss below. On the other hand, when $\frac{\partial h_A}{\partial R_B} \frac{\partial h_B}{\partial R_A} < 0$, the eigenvalues are purely imaginary, corresponding to an instability that develops as a spiral motion around the fixed point. The fixed point loses stability at $\Lambda_{\pm}=\pm i$, which is a special case of Neimark--Sacker bifurcation \cite{kuznetsov1998elements}. The eigenvalues $\pm i$ correspond to a rotation of $\pi/2$ hence this scenario leads to a period-4 oscillation (see Figure \ref{fig:d_p2_p4} (i-l)). 
Interestingly, when the triadic regulations are exclusively interlayer, even if the regulations are exclusively positive, it is still possible to observe a period-two oscillation when $\Lambda_+=1$ (see Figure \ref{fig:period2}). Thus, this scenario is very different from the case of triadic percolation on single-layer networks, in which periodic oscillation can be observed only in the presence of negative regulations.
To see this, let us decouple the two-dimensional map by considering the iterative maps, 
\bea
R_A^{(t)} &=& h_A\left(h_B(R_A^{(t-2)};p); p\right) \nonumber\\
R_B^{(t)} &=& h_B\left(h_A(R_B^{(t-2)};p); p\right)
\label{eq:second_iteration_map}
\eea
valid for \revise{$t \geq 2$} having initial conditions $R_{A/B}^{(1)}$ for odd times and initial conditions $R_{A/B}^{(2)}$ for even times. These initial conditions are all determined by the initial probabilities $p_A^{(0)}$ and $p_B^{(0)}$ to retain the links in layer A and layer B. Indeed we have 
\bea
R_A^{(1)} &=& f_A\left(p_{A}^{(0)}\right), \quad R_B^{(1)} = f_B\left(p_B^{(0)}\right); \nonumber\\
R_A^{(2)} &=& h_A\left(f_B\left(p_B^{(0)}\right)\right), \quad R_B^{(2)} = h_B\left(f_a\left(p_A^{(0)}\right)\right).
\label{eq:related_initial_condition}
\eea
\revise{Hence, the dynamics of even and odd times are completely independent and} the full dynamics of $R_A^{(t)}$ and $R_B^{(t)}$ can be obtained by integrating either one of the two one-dimensional maps in Eq.~(\ref{eq:second_iteration_map}). For example one can consider the one-dimensional maps determining the even and odd times of the timeseries $R_A^{(t)}$ having initial conditions $R_A^{(1)}$ and $R_A^{(2)}$ respectively. Once the full time series of $R_A^{(t)}$ is determined by interleaving the solution obtain for even and odd times,  the timeseries for  $R_B^{(t)}$ can be determined using the second of Eqs.~(\ref{map_inter}).

The one-dimensional maps  $R_A^{(t)}=h_A\left(h_B(R_A^{(t-2)};p); p\right)$ corresponding to even and odd times are identical, however they have different initial conditions. This implies that if the map has more than one stable fixed point, the even and the odd timeseries might, under suitable initial conditions, converge to two different fixed points.
For example this is what can occur if the interlayer triadic interactions are exclusively positive, a situation in which we can observe two stable fixed points one corresponding to $R_A^{\star}=0$ and one corresponding to $R_A^{\star}=\bar{R}^{\star}>0$ in addition to one unstable fixed point.  The non-trivial stable fixed point $\bar{R}^\star>0$ emerges at $\left.\frac{\partial h_A(h_B(R))}{\partial R}\right|_{R=\bar{R}^\star}=1$, i.e., ${\Lambda}=1$. (see Figure \ref{fig:period2} (a)).  Under this setting, depending on the initial conditions we can have three possible scenarios:
\begin{itemize}
    \item 1. Both even and odd timeseries converge to $R^{\star}=0$, hence the full dynamics converges to $R^{\star}=0$ (see Figure \ref{fig:period2} (b));
    \item 2. Both even and odd timeseries converge to $\bar{R}^{\star}$, hence the full dynamics converges to $\bar{R}^{\star}>0$ (see Figure \ref{fig:period2} (c));
    \item 3. One timeseries converges to 
    $R^{\star}=0$ and the other converges to $\bar{R}^{\star}$, hence the full dynamics displays a period-two oscillation between $R^{\star}=0$ and $\bar{R}^{\star}>0$ (see Figure \ref{fig:period2} (d)).
\end{itemize}
The third scenario above indicates that in MTP with exclusively positive interlayer triadic interactions we can observe period-two oscillations of $R_A^{(t)}$. By considering the timeseries of $R_B^{(t)}$ in a number of cases we can also observe period-two oscillations
(see Figure $\ref{fig:period2}$). This is an interesting difference from the triadic percolation on single-layer networks studied in Ref. \cite{sun2023dynamic}, where only discontinuous transitions can be observed if the triadic interactions are exclusively positive. 
Interestingly, due to the dependency of initial conditions, the phase diagram may display a discontinuous hybrid transition (see Figure \ref{fig:period2} (f)) or a discontinuous jump from period-two oscillation to zero (Figure \ref{fig:period2} (e)).

When the considered one-dimensional map $R_A^{(t)}=h_A\left(h_B(R_A^{(t-2)};p); p\right)$ undergoes a period-doubling bifurcation using similar arguments, it is possible to show that for suitable initial conditions, a period-4 oscillation emerges as a result of combining two period-two oscillations. This corresponds to the Neimark--Sacker bifurcation of the 2-dimensional map mentioned before. 

Note that the above argument, developed starting from the one-dimensional map $R_A^{(t)}=h_A\left(h_B(R_A^{(t-2)};p); p\right),$ can be carried out analogously considering instead the map $R_B^{(t)}=h_B\left(h_A(R_B^{(t-2)};p); p\right)$.

\section{Characterization of the phase diagram }

Here we aim to provide a more complete description of the critical properties of MTP  and investigate how they are affected by the multilayer network structures. We focus on the general triadic percolation with both inter- and intra-layer regulations, highlight the stability of fixed points, and study where/how they lose stability as one varies the control parameter $p$. Specifically, our analysis focuses on the three critical thresholds, namely the upper stability threshold $p_c^u$, the re-stabilization threshold $p_c^s$, and the lower stability threshold $p_c^l$ that are illustrated in  Figure \ref{fig:orbit} (b). At the upper threshold $p_c^u$, the non-trivial fixed point first loses stability; at the re-stabilization threshold $p_c^s$, a stable fixed point re-emerges after chaotic or oscillatory dynamics; and at the lower threshold $p_c^l$, this non-trivial fixed point destabilizes again, leaving the trivial zero state as the only stable fixed point. Note that, unlike the single-layer case where instability typically arises through period-doubling, in multilayer triadic percolation the fixed point may also lose stability via a Neimark–Sacker bifurcation, leading to a richer variety of orbit diagrams.


Here, we study the three critical thresholds as a function of the strength of intra-layer and inter-layer regulations, namely $c_{A_\text{intra}}^{-}$ and  $c_{A_\text{inter}}^{-}$. We calculate the critical values and characterize the natures of the bifurcations. Note that we limit the discussion on the bifurcation structure of the fixed points in parameter space, i.e., how fixed points lose stability through various critical bifurcations, without addressing the full orbit diagram or the route to chaos. However, we observe that the occurrence of the Neimark–Sacker might be connected with the quasi-periodic route to chaos also known as Ruelle–Takens–Newhouse scenario. The investigation of whether the route to chaos of MTP really follows in this universality class is beyond the scope of this work and will be addressed in subsequent publications.

\begin{figure*}
    \centering
    \includegraphics[width=\linewidth]{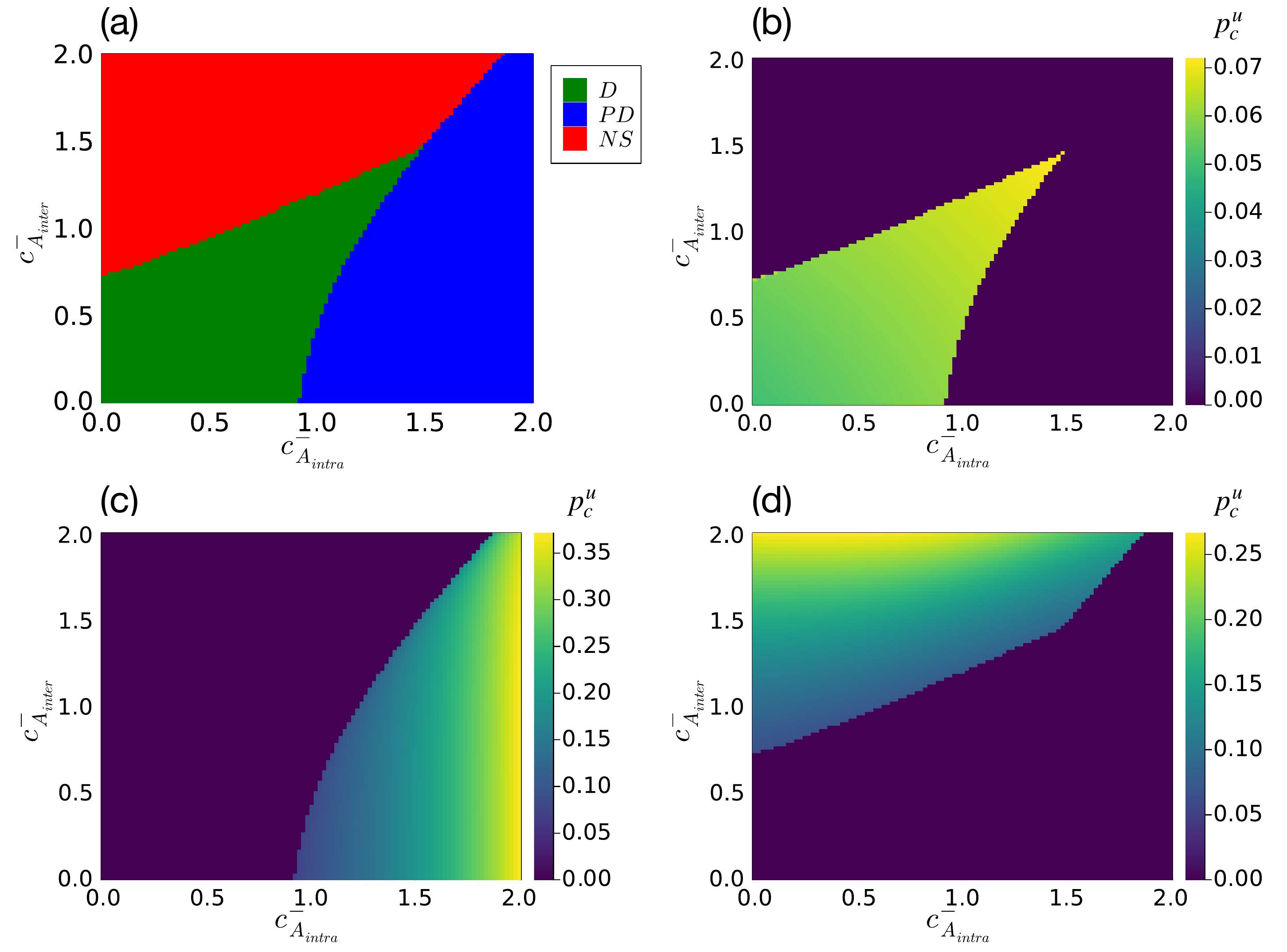}
    \caption{Phase diagram of the upper stability threshold $p_c^u$. Panel (a) shows the nature of the phase transition, discontinuous ($D$, green), period-doubling bifurcation ($PD$, blue) and Neimark--Sacker bifurcation ($NS$, red) under different model parameters $c_{A_\text{intra}}^{-}$ and  $c_{A_\text{inter}}^{-}$. The heatmaps in panel (b), (c) and (d) show the critical $p_c^u$ value of discontinuous transition (b), period-doubling transition (c) and Neimark--Sacker transition (d).
    The upper stability threshold $p_c^u$ denotes the largest value of $p$ where the non-zero fixed point loses stability.
    Poisson structural networks with averaged degrees $c_A=c_B=30$ are considered in the numerical calculations. The intralayer and interlayer regulatory networks are also Poissonian. The fixed parameters are $c_{A_\text{intra}}^{+} = 10$, $c_{A_\text{inter}}^{+} = 10$, $c_{B_\text{intra}}^{+} = \infty$, $c_{B_\text{intra}}^{-} = 0$, $c_{B_\text{inter}}^{+} = 3$, $c_{B_\text{inter}}^{-} = 0$. }
    \label{fig:bifurcation_phase_diagram}
\end{figure*}

\begin{figure*}
    \centering
    \includegraphics[width=\linewidth]{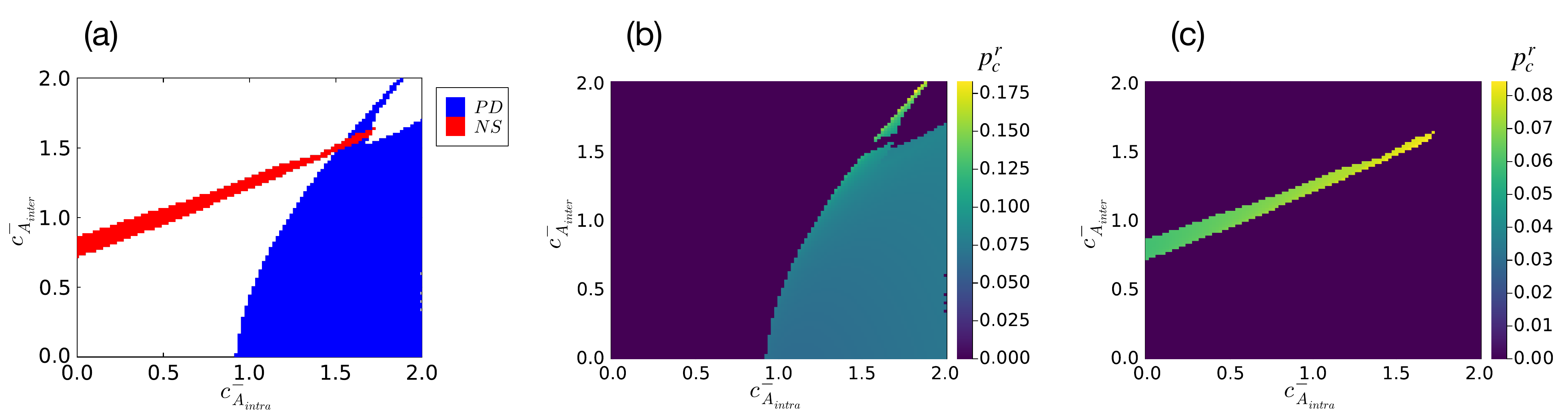}
    \caption{Phase diagram of the re-stabilization threshold $p_c^s$. Panel (a) shows the nature of the phase transition of period-doubling bifurcation ($PD$, blue) and Neimark--Sacker bifurcation ($NS$, red) under different model parameters $c_{A_\text{intra}}^{-}$ and  $c_{A_\text{inter}}^{-}$. Panels (b) and (c) show the critical $p_c^s$ value of the period-doubling transition (b) and Neimark--Sacker transition (c). The smallest bifurcation point $p_c^l$ denotes the smallest $p$ value at which the non-zero fixed point loses stability. Poisson structural networks with averaged degrees $c_A=c_B=30$ are considered in the numerical calculations. The intralayer and interlayer regulatory networks are also Poissonian. The fixed parameters are $c_{A_\text{intra}}^{+} = 10$, $c_{A_\text{inter}}^{+} = 10$, $c_{B_\text{intra}}^{+} = \infty$, $c_{B_\text{intra}}^{-} = 0$, $c_{B_\text{inter}}^{+} = 3$, $c_{B_\text{inter}}^{-} = 0$.}
    \label{fig:bifurcation_phase_diagram_left}
\end{figure*}

\begin{figure*}
    \centering
    \includegraphics[width=\linewidth]{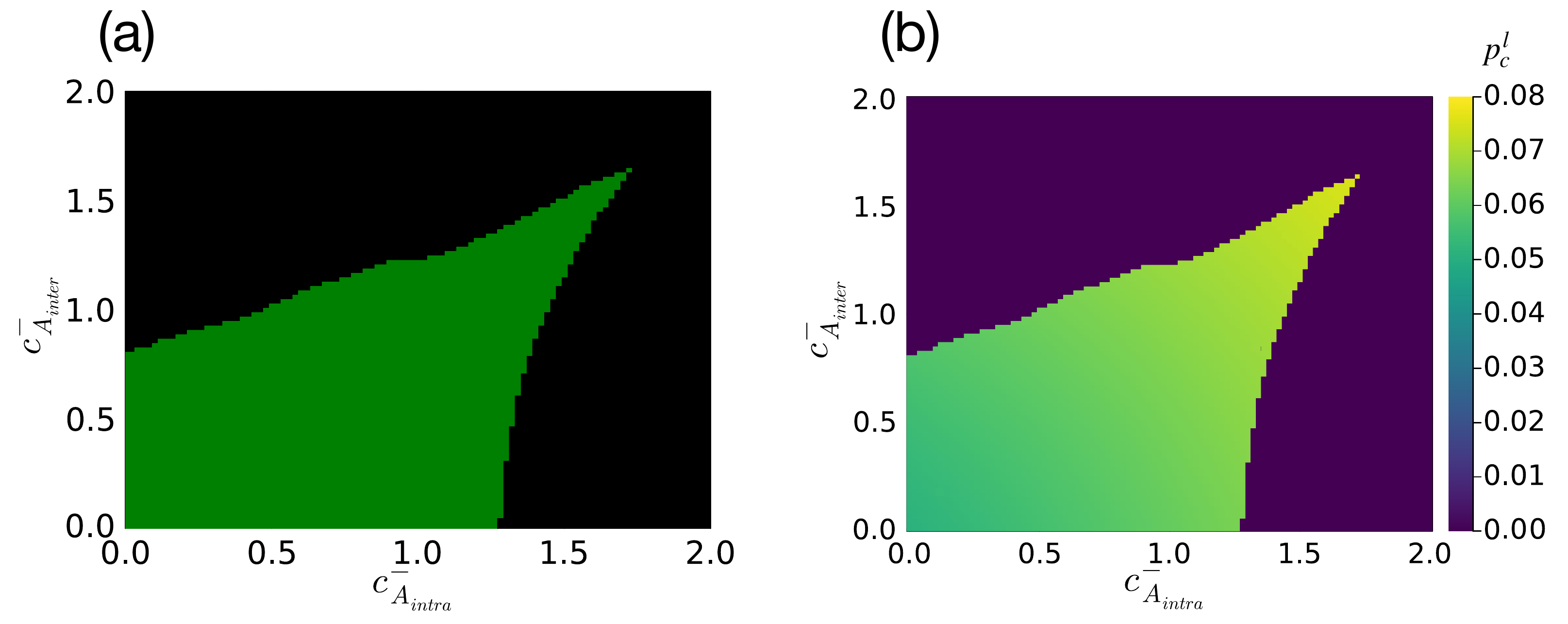}
    \caption{Phase diagram of the lower stability threshold $p_c^l$. Panel (a) shows the nature of the phase transition under different model parameters $c_{A_\text{intra}}^{-}$ and $c_{A_\text{inter}}^{-}$. The green region denotes the discontinuous transition from a nontrivial steady state to zero, and the black region denotes the abrupt transition from periodic cycles or chaos to zero via attractor destruction. Panel (b) shows the critical $p_c^l$ value of the discontinuous transition as a function of $c_{A_\text{intra}}^{-}$ and $c_{A_\text{inter}}^{-}$. The lower stability threshold $p_c^l$ denotes the critical point under which only the trivial zero fixed point is stable.
    Poisson structural networks with averaged degrees $c_A=c_B=30$ are considered in the numerical calculations. The intralayer and interlayer regulatory networks are also Poissonian. The fixed parameters are $c_{A_\text{intra}}^{+} = 10$, $c_{A_\text{inter}}^{+} = 10$, $c_{B_\text{intra}}^{+} = \infty$, $c_{B_\text{intra}}^{-} = 0$, $c_{B_\text{inter}}^{+} = 3$, $c_{B_\text{inter}}^{-} = 0$.}
    \label{fig:bifurcation_phase_diagram_from_zero}
\end{figure*}

In Figures \ref{fig:bifurcation_phase_diagram}, \ref{fig:bifurcation_phase_diagram_left}, and \ref{fig:bifurcation_phase_diagram_from_zero}, we show the critical points $p_c^u$, $p_c^s$ and $p_c^l$ respectively as functions of the model parameters $c_{A_{\text{intra}}}^-$ and $c_{A_{\text{inter}}}^-$. At the upper threshold $p_c^u$, the non-trivial fixed point loses stability. Depending on the parameter values, this destabilization may occur via a discontinuous, a period-doubling, or a Neimark--Sacker bifurcation. Interestingly, due to the change of the nature of the phase transition while changing the model parameter, the critical point $p_c^u$ displays a highly non-trivial non-monotonic relationship with the model parameter. This indicates that under certain circumstances, increasing the negative regulation strength could stabilize the steady state (see Figure \ref{fig:bifurcation_phase_diagram} (b-d)). This phenomenon was not observed on the single-layer triadic percolation model. 

When the transition at $p_c^u$ is discontinuous, the trivial fixed point remains the only stable fixed point of the system. Hence at the re-stabilization threshold $p_c^s$, the trivial fixed point regains stability, but only through period-doubling or Neimark--Sacker bifurcations. Finally, at the lower stability threshold $p_c^l$, the non-trivial fixed point typically destabilizes via a discontinuous transition. However, for certain parameter regimes, the system may undergo an abrupt transition from oscillatory or chaotic dynamics to the trivial steady state due to attractor destruction, i.e., when the trajectory collides with the basin boundary.

These analyses illustrate that the nature of the bifurcation depends sensitively on the balance between intra-layer and inter-layer regulations, resulting in qualitatively and quantitatively different dynamical behaviors. Hence, we reveal the complex interplay between network structure and bifurcation scenarios, and we highlight the non-trivial co-existence of multiple bifurcation types within a given network configuration.

\section{Conclusion}\label{sec:conclusion}

In this study, we propose a comprehensive framework of multilayer triadic percolation. In multilayer triadic percolation, links can be up- or down-regulated by triadic interactions, turning percolation into a fully-fledged dynamical process. Incorporating percolation theory with non-linear dynamics, we reveal that this dynamical process induced by intra- and inter-layer triadic regulatory interactions can account for time-varying giant components on each layer of the multilayer network.  Inherent to the multilayer structure, triadic percolation on multilayer networks exhibits richer critical phenomena than in single-layer networks. Here we have focused on the properties of the bifurcation transitions in multilayer triadic percolation and we have shown that in the thermodynamic limit, the system can undergo a Neimark–Sacker bifurcation, which arises uniquely in multilayer triadic percolation and has no analogue in the single-layer counterpart. Moreover, we find that period-doubling oscillations may also occur under exclusively positive regulations, giving rise to dynamical states that are absent in single-layer systems.


In addition, we explore the complex interplay between network structure and the critical phenomena of multilayer triadic percolation. The analyses are restricted to the bifurcation of fixed points in parameter space. The characterisation of the full orbit diagram and the route to chaos will be left for future work. We reveal that both the types of bifurcations and the corresponding critical values depend sensitively on the interplay between intra- and inter-layer regulatory interactions. The results highlight that these regulations can generate not only quantitative shifts in the stability thresholds but also qualitative differences in the dynamical behaviors. 

\revise{ Here we have focused specifically on MTP defined on multilayer networks with $M=2$ layers. As discussed briefly in this work, extending our results to multilayers with $M>2$ will involve treating MTP using higher-dimensional maps of dimension $M$. While we acknowledge that the resulting dynamical process, and specifically the route to chaos, might imply complex dynamical behavior not captured by the $M=2$ maps, we observe that the bifurcation mechanisms which have been our focus in this manuscript, can be more directly generalizable to multilayer networks with $M>2$. A systematic exploration of the possible additional phenomenology arising for a higher dimensional maps induced by the multilayer network structure is beyond the scope of the present paper and will be addressed in subsequent works.}

The multilayer framework provides a more realistic representation of networks with triadic regulatory interactions. Within this comprehensive framework, we uncover the rich interplay between multilayer network structure and critical phenomena of triadic percolation, offering new insights into the dynamics of real-world systems, including brain and ecological networks. In addition, due to the rich dynamical behavior of multilayer triadic percolation, the proposed framework might provide a promising tool for adaptively controlling network activity.



\begin{acknowledgements}
H.S., F.R. and G.B. acknowledge support by AccelNet-MultiNet program, a project of the National Science Foundation (Award \#1927425 and \#1927418).
 H.S. acknowledges the support of the Wallenberg Initiative on Networks and Quantum Information (WINQ). F.R. acknowledges support from the Air Force Ofﬁce of Scientiﬁc Research under grant number FA9550-24-1-0039.
The funders had no role
in study design, data collection, and analysis, the decision to publish, or
any opinions, ﬁndings, conclusions, or recommendations expressed in
the manuscript. H.S. thanks Zhuan Li and Joaquín J. Torres for the inspiring discussions on the initial stage of this project.
\end{acknowledgements}

\bibliographystyle{unsrt}
\bibliography{reference}

\end{document}